\pdfoutput=1
\documentclass[aps,prd,preprint,superscriptaddress,tightenlines,nofootinbib]{revtex4}

\long\def\symbolfootnote[#1]#2{\begingroup%
\def\thefootnote{\fnsymbol{footnote}}\footnote[#1]{#2}\endgroup}

\usepackage{amsfonts,amsmath,amsthm,amssymb}
\usepackage{mathrsfs}
\usepackage{bm}
\usepackage{color}
\usepackage{epsfig}

\newcommand{\PRE}[1]{{#1}}   

\newcommand{\beq}{\begin{equation}}
\newcommand{\eeq}{\end{equation}}
\newcommand{\bea}{\begin{flushleft} \begin{eqnarray}}
\newcommand{\eea}{\end{eqnarray}\end{flushleft}}

\newcommand{\postscript}[2]{\setlength{\epsfxsize}{#2\hsize}
   \centerline{\epsfbox{#1}}}
\newcommand{\comment}[1]{}

\newcommand{\ci}[1]{}

\newcommand{\ba}{\begin{eqnarray}}
\newcommand{\ea}{\end{eqnarray}}
\newcommand{\be}{\begin{equation}}
\newcommand{\ee}{\end{equation}}
\newcommand{\bay}[1]{\left(\begin{array}{#1}}
\newcommand{\eay}{\end{array}\right)}

\def\xt{{\theta}}

\def\CD{{\cal D}}

\begin{document}

\preprint{
\hfil
\begin{minipage}[t]{3in}
\begin{flushright}
\vspace*{.4in}
MPP--2012--124\\
LMU-ASC 58/12\\
CERN-PH-TH/2012-222\\
\end{flushright}
\end{minipage}
}

\title{\PRE{\vspace*{0.9in}}
Vacuum Stability of Standard Model$\bm{^{++}}$}

\author{{\bf Luis A. Anchordoqui}}
\affiliation{Department of Physics,\\
University of Wisconsin-Milwaukee,
 Milwaukee, WI 53201, USA
\PRE{\vspace*{.05in}}
}

\author{{\bf Ignatios Antoniadis}}\thanks{On leave of absence
from CPHT Ecole Polytechnique, F-91128, Palaiseau Cedex.}
\affiliation{Department of Physics,\\ CERN Theory Division,
CH-1211 Geneva 23, Switzerland
\PRE{\vspace*{.05in}}}

\author{{\bf Haim \nolinebreak Goldberg}}
\affiliation{Department of Physics,\\
Northeastern University, Boston, MA 02115, USA
\PRE{\vspace*{.05in}}
}

\author{{\bf Xing Huang}}
\affiliation{School of Physics and Astronomy, \\
Seoul National University, Seoul 141-747, Korea
\PRE{\vspace*{.05in}}
}

\author{{\bf Dieter L\"ust}}

\affiliation{Max--Planck--Institut f\"ur Physik, \\ 
 Werner--Heisenberg--Institut,
80805 M\"unchen, Germany
\PRE{\vspace*{.05in}}
}

\affiliation{Arnold Sommerfeld Center for Theoretical Physics 
Ludwig-Maximilians-Universit\"at M\"unchen,
80333 M\"unchen, Germany
\PRE{\vspace{.05in}}
}

\author{{\bf Tomasz R. Taylor}}

\affiliation{Department of Physics,\\
 Northeastern University, Boston, MA 02115, USA 
 \PRE{\vspace*{.05in}}
}

\author{{\bf Brian Vlcek}}
\affiliation{Department of Physics,\\
University of Wisconsin-Milwaukee,
 Milwaukee, WI 53201, USA
\PRE{\vspace*{.05in}}
}


\PRE{\vspace*{.15in}}

\begin{abstract}\vskip 2mm
  \noindent 
  The latest results of the ATLAS and CMS experiments point to a
  preferred narrow Higgs mass range ($m_H \simeq 124 - 126~{\rm GeV}$)
  in which the effective potential of the Standard Model (SM) develops
  a vacuum instability at a scale $10^{9} - 10^{11}~{\rm GeV}$, with
  the precise scale depending on the precise value of the top quark
  mass and the strong coupling constant.  Motivated by this
  experimental situation, we present here a detailed investigation
  about the stability of the SM$^{++}$ vacuum, which is characterized
  by a simple extension of the SM obtained by adding to the scalar
  sector a complex $SU(2)$ singlet that has the quantum numbers of the
  right-handed neutrino, $H''$, and to the gauge sector an $U(1)$ that
  is broken by the vacuum expectation value of $H''$.  We derive the
  complete set of renormalization group equations at one loop.  We
  then pursue a numerical study of the system to determine the
  triviality and vacuum stability bounds, using a scan of $10^4$
  random set of points to fix the initial conditions.  We show that,
  if there is no mixing in the scalar sector, the top Yukawa coupling
  drives the quartic Higgs coupling to negative values in the
  ultraviolet and, as for the SM, the effective potential develops an
  instability below the Planck scale. However, for a mixing angle
  $-0.35 \alt \alpha \alt -0.02$ or $0.01 \alt \alpha \alt 0.35$, with
  the new scalar mass in the range $500~{\rm GeV} \alt m_{h''} \alt
  8~{\rm TeV}$, the SM$^{++}$ ground state can be absolutely stable up
  to the Planck scale.  These results are largely independent of
  TeV-scale free parameters in the model: the mass of the
  non-anomalous $U(1)$ gauge boson and its branching fractions.

  \end{abstract}

\maketitle

\tableofcontents

\section{Introduction}

The CERN Large Hadron Collider (LHC) has begun a bound and determined
exploration of the electroweak scale. Recently, the
ATLAS~\cite{Gianotti} and CMS~\cite{Incandela} collaborations
presented an update of the Higgs searches, independently combining
about $5~{\rm fb}^{-1}$ of data collected at $\sqrt{s} = 7~{\rm TeV}$
and more than $5~{\rm fb}^{-1}$ at $\sqrt{s} = 8~{\rm TeV}$. The
excess at 125~GeV that was evident already in data from the 7~TeV
run~\cite{ATLAS:2012ae,Chatrchyan:2012tx} has been consistently
observed by both experiments in the $\gamma \gamma$ invariant mass
spectrum with a local significance of $4.5\sigma$ and $4.1\sigma$,
respectively. In addition, an excess of 4 leptons events (with
$m_{4\ell} \simeq 125~{\rm GeV}$) which can be interpreted as a signal
of the $H \to ZZ^* \to 4 \ell$ decay, is observed by both experiments
with a significance of $3.4\sigma$ and $3.2\sigma$, respectively. The
CMS experiment also presented updated Higss boson searches in $W^+W^-$
(a broad excess in the invariant mass distribution of $1.5\sigma$ is
observed), $b\bar b$ (no excess is observed), and $\tau \bar \tau$ (no
excess is observed) channels.  More recently, the ATLAS Collaboration
reported a $2.8\sigma$ deviation in the $H \to W^+ W^- \to 2 \ell \nu$
decay channel~\cite{Arnaez}.  When combining the data from the 7~TeV
and 8~TeV runs, both experiments separately have reached the
sensitivity to the new boson with a local significance of
$5\sigma$~\cite{ATLASnew,CMSnew}.  Very recently, the CDF and
D0 collaborations published an update on searches for the Higgs boson
decaying into $b \bar b$ pairs, using $9.7~{\rm fb}^{-1}$ of data
collected at $\sqrt{s} = 1.96~{\rm TeV}$~\cite{Aaltonen:2012qt}. They
reported a $3.3\sigma$ deviation with respect to the background-only
hypothesis in the mass range between $120 - 135~{\rm GeV}$.

All in all, LHC data strongly suggest that the observed state feeds the
electroweak symmetry breaking, and is likely the Higgs boson. However,
it remains to be seen whether its trademarks, particularly the
production cross section and decay branching fractions, agree with the
very precise prediction of the Standard Model (SM). The decay channels
that are most sensitive to new physics are the loop induced decays $H 
\to \gamma \gamma$ and $H \to \gamma Z$. Interestingly, the most
recent analyses~\cite{Giardino:2012dp,Espinosa:2012im} of the combined
LHC data seem to indicate that there is a deviation from SM
expectations in the diphoton channel at the $2.0\sigma - 2.3\sigma$
level; see, however,~\cite{Plehn:2012iz}.

From a theoretical perspective some modification of the Higgs sector 
has long been expected, since the major motivation for physics beyond
the SM is aimed at resolving the hierarchy problem. Even if one
abandons such a motivation for new physics there are still enduring
concerns about the stability of the electroweak vacuum, which have
been exacerbated by the new LHC  data that points to $m_H
\simeq 125~{\rm GeV}$.

Next-to-leading order (NLO) constraints on SM vacuum stability based
on two-loop renormalization group (RG) equations, one-loop threshold
corrections at the electroweak scale (possibly improved with two-loop
terms in the case of pure QCD corrections), and one-loop effective
potential seem to indicate $m_H \approx 125 - 126~{\rm GeV}$ saturates
the minimum value that ensures a vanishing Higgs quartic coupling
around the Planck scale ($M_{\rm Pl}$), see {\it
  e.g.}~\cite{Lindner:1988ww,Sher:1988mj,Diaz:1994bv,Casas:1994qy,Diaz:1995yv,Casas:1996aq,Isidori:2001bm,Isidori:2007vm,Hall:2009nd,Ellis:2009tp,EliasMiro:2011aa,Xing:2011aa}. However,
the devil is in the details, a more recent NNLO
analysis~\cite{Bezrukov:2012sa,Degrassi:2012ry} yields a very
restrictive condition of absolute stability up to the Planck scale
\begin{equation}
m_H> \left[129.4 + 1.4 \left( \frac{m_t/{\rm GeV} -173.1}{0.7}
\right) - 0.5 \left(\frac{\alpha_s(m_Z) - 0.1184}{0.0007} \right) \pm
1.0_{\rm th} \right]~{\rm GeV}  \, .
\label{1}
\end{equation}
When combining in quadrature the theoretical uncertainty with
experimental errors on the mass of the top ($m_t$) and the strong
coupling constant ($\alpha_s$), one obtains $m_H > 129 \pm 1.8~{\rm
  GeV}$. The vacuum stability of the SM up to the Planck scale is
excluded at 2$\sigma$ (98\% C.L. one sided) for $m_H < 126~{\rm
  GeV}$~\cite{Bezrukov:2012sa,Degrassi:2012ry}. Achieving the
stability will necessarily impose constrains on physics beyond the SM.

Very recently we have put forward a (string based) Standard-like
Model~\cite{Anchordoqui:2012wt}. Motivated by the above, here we study
the vacuum stability of its scalar sector.  The layout of the paper is
as follows.  In Sec.~\ref{II} we briefly review the generalities of
our model and we derive the RG equations. In Sec.~\ref{III} we present
our results and conclusions.

\section{RG Evolution Equations of SM$\bm{^{++}}$}
\label{II}

\begin{figure}[tbp]
\postscript{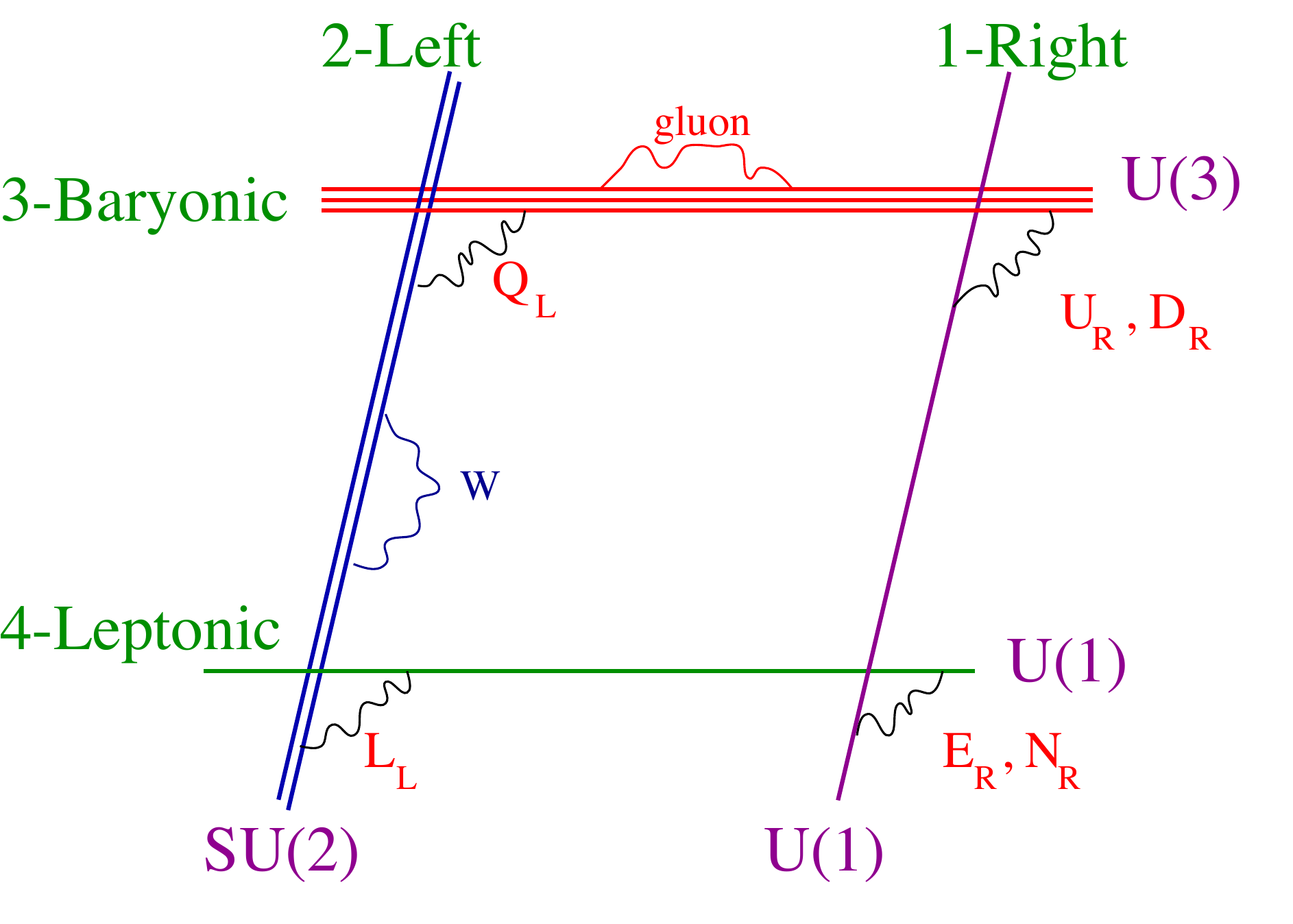}{0.8}
\caption{Pictorial representation of the $U(3)_B \times SU(2)_L \times U(1)_L \times U(1)_{I_R}$ D-brane model.}
\label{cartoon}
\end{figure}

Very recently, we engineered the minimal extension of the SM that can
be embedded into a Superstring Theory endowed with a high mass string
scale, $M_s \alt M_{\rm Pl}$~\cite{Anchordoqui:2012wt}. The gauge
extended sector, $U(3)_B \times SU(2)_L \times U(1)_{I_R} \times
U(1)_L$, has two additional $U(1)$ symmetries and thus we refer to our
model as SM$^{++}$. The origin of this model is founded on the D-brane
structure of string compactifications, with all six extra dimensions
${\cal O} (M_{\rm
  Pl}^{-1})$~\cite{Lust:2004ks,Blumenhagen:2005mu,Blumenhagen:2006ci}. The
low energy remnants of the D-brane structure are the gauge bosons and
Weyl fermions living at the brane intersections of a particular 4-stack
quiver configuration~\cite{Cremades:2003qj}. A schematic representation of the
D-brane construct is shown in Fig.~\ref{cartoon}.  The general
properties of the chiral spectrum are summarized in Table~\ref{table}.

The resulting $U(1)$ content gauges the baryon number $B$ [with
$U(1)_B \subset U(3)_B$], the lepton number $L$, and a third
additional abelian charge $I_R$ which acts as the third isospin
component of an $SU(2)_R$.  Contact with gauge structures at TeV
energies is achieved by a field rotation to couple diagonally to
hypercharge $Y_\mu$. Two of the Euler angles are determined by this
rotation and the third one is chosen so that one of the $U(1)$ gauge
bosons couples only to an anomaly free linear combination of $I_R$ and
$B-L$.  Of the three original abelian couplings, the baryon number
coupling $g'_3$ is fixed to be $\sqrt{1/6}$ of the QCD coupling $g_3$
at the string scale. The orthogonal nature of the rotation imposes one
additional constraint on the remaining couplings $g'_1$ and
$g'_4$~\cite{Anchordoqui:2011eg}.  Since one of the two extra gauge
bosons is coupled to an anomalous current, its mass is ${\cal
  O}(M_s)$, as generated through some St\"uckelberg
mechanism.\footnote{A point worth noting at this juncture: SM can also
  be embedded in a 3-stack quiver comprising (only) one additional
  $U(1)$ symmetry, $U(3) \times SU(2) \times
  U(1)$~\cite{Berenstein:2006pk}. The extra gauge boson is anomalous
  and must grow a St\"uckelberg mass $\sim M_s$. In this D-brane model
  the running of the quartic Higgs coupling would reveal a vacuum
  instability around $10^{10}~{\rm
    GeV}$~\cite{Bezrukov:2012sa,Degrassi:2012ry}.}  The other gauge
boson is coupled to an anomaly free current and therefore (under
certain topological conditions) it can remain massless and grow a
TeV-scale mass through ordinary Higgs mechanisms~\cite{Cvetic:2011iq}.

Electroweak symmetry breaking is achieved through the standard Higgs
doublet $H$. The spontaneous symmetry breaking of the extra
non-anomalous $U(1)$ is attained through an $SU(2)$ singlet scalar
field $H''$, which carries $L$ and $I_R$ numbers, and acquires a
vacuum expectation value (VEV) at the TeV scale.  With the charge
assignments of Table~\ref{table} there are no dimension 4 operators
involving $H''$ that contribute to the Yukawa Lagrangian. This is very
important since $H''$ carries the quantum numbers of right-handed
neutrino and its VEV breaks lepton number. However, this breaking can
affect only higher-dimensional operators which are suppressed by the
high string scale, and thus there is no phenomenological problem with
experimental constraints for $M_s$ higher than $\sim 10^{14}$
GeV. Herein we remain agnostic with respect to supersymmetry (SUSY)
breaking and the the details of the low energy effective
potential. However, we do subject the choice of quantum numbers for
$H''$ to the stringent holonomic constraints of the superpotential at
the string scale.  This forbids the simultaneous presence of scalar
fields and their complex conjugate. As an illustration, if the quantum
numbers of $H''$ are those of $N_R^c$, then higher dimensional
operators such as $\overline N_R N_R^c {H''}^2$, which can potentially
generate a Majorana mass, are absent.\footnote{The identification of $H''$ with the superpartner of the right handed neutrino has been noted in~\cite{Kadastik:2009cu}.} Because of holonomy this absence
cannot be circumvented by including $\overline N_R N_R^c {H''}^{*2}$.

\begin{table}
  \caption{Chiral  spectrum of SM$^{++}$.}
\begin{center}
\begin{tabular}{ccccccc}
\hline
\hline
~~~Fields~~~ & ~~~Sector~~~  & ~~~Representation~~~ & ~~~$Q_B$~~~ & ~~~$Q_L$~~~ & ~~~$Q_{I_R}$~~~ & ~~~$Q_Y$~~~ \\
\hline
 $U_R$ &   $\phantom{^*}3 \leftrightharpoons 1^*$ &  $(3,1)$ & $1$ & $\phantom{-}0$ & $\phantom{-} 1$ & $\phantom{-}\frac{2}{3}$  \\[1mm]
  $D_R$ & $3 \leftrightharpoons 1$  & $( 3,1)$&    $1$ & $\phantom{-}0$ & $- 1$ & $-\frac{1}{3}$   \\[1mm]
  $L_L$ & $4 \leftrightharpoons 2$ &  $(1,2)$&    $0$ &  $\phantom{-}1$ & $\phantom{-}0$ & $-\frac{1}{2}$ \\[1mm]
  $E_R$ & $4 \leftrightharpoons 1$ &  $(1,1)$&   $0$ & $\phantom{-}1$ &  $- 1$ & $- 1$ \\[1mm]
 $Q_L$ & $3 \leftrightharpoons 2$ &  $(3,2)$& $1$ & $\phantom{-}0 $ & $\phantom{-} 0$ & $\phantom{-} \frac{1}{6}$    \\[1mm]
   $N_R$  &  $\phantom{^*}4 \leftrightharpoons 1^*$  &   $(1,1)$& $0$ & $\phantom{-}1$ & 
$\phantom{-} 1$ & $\phantom{-} 0$ \\ [1mm]
$H$ & $2 \leftrightharpoons 1$ &  $(1,2)$ & $0$ & $\phantom{-}0$ & $\phantom{-}1 $ &
$\phantom{-}\frac{1}{2}$  \\ [1mm]
$H''$ & $4 \leftrightharpoons 1$ &   $(1,1)$ & $0$ & $-1$ & $-1$  & $\phantom{-}0$ \\ 
\hline
\hline
\label{table}
\end{tabular}
\end{center}
\end{table}

The scalar Lagrangian of SM$^{++}$ is
\begin{equation}\label{new-scalar_L}
\mathscr{L}_s^{++} =\left( {\cal D}^{\mu} H\right) ^{\dagger} {\cal D}_{\mu}H + 
\left( {\cal D}^{\mu} H'' \right) ^{\dagger} {\cal D}_{\mu}H'' - V^{++}(H,H'' ) \, ,
\end{equation}
where 
\beq
V^{++} \left(H, H'' \right) = \mu_1^2 \left| H \right|^2 +{ \mu_2}^2 \left| H''
\right|^2 + \lambda_1 \left| H \right|^4 + \lambda_2 \left| H''
\right|^4 + \lambda_3 \left| H \right|^2 \left| H'' \right|^2 
\label{higgsV}
\eeq 
is the potential and 
\beq {\cal D}_\mu = \partial_\mu - i g_3 T^a
G^a_\mu - i g'_3 Q_B C_\mu - i g_2 \tau^a W^a_\mu - i g'_1 Q_{I_R}
B_\mu - i g'_4 Q_{L} X_\mu
\label{covderi2}
\eeq is (in a self-explanatory notation~\cite{Anchordoqui:2012wt}) the
covariant derivative in the field basis shown in Fig.~\ref{cartoon}.
Next, we impose the positivity conditions~\cite{Barger:2008jx}
\begin{equation}
\lambda_1 > 0, \quad \quad \lambda_2 > 0, \quad \quad  \lambda_1
\lambda_2 > \frac{1}{4} \lambda_3^2 \, .
\label{VernonGabePaul}
\end{equation}
If the conditions (\ref{VernonGabePaul}) are satisfied, we can proceed
to the minimisation of $V^{++} (H, H'')$ as a function of constant VEVs for
the two Higgs fields. In the unitary gauge the fields can be written as
\begin{equation}\label{min}
 H  \equiv \frac{1}{\sqrt{2}} 
  \left( \begin{array}{c} 0 \\ v_1 + h_1(x)\end{array} \right)
\quad {\rm and} \quad  H''  \equiv \frac{1}{\sqrt{2}} \left(
v_2 + h_2(x) \right)\, ,
\end{equation} 
with $v_1$ and $v_2$ real and non-negative. The physically most
interesting solutions to the minimisation of (\ref{higgsV}) are
obtained for $v_1$ and $v_2$ both non-vanishing
\begin{equation}
v_1^2 = \frac{-\lambda _2 \mu_1^2 + \frac{1}{2} \lambda _3 {\mu_2}
  ^2}{\lambda _1 \lambda _2 - \frac{1}{4} \lambda _3^2}
\quad {\rm and} \quad
v_2^2 = \frac{-\lambda _1 \mu_2^2 + \frac{1}{2} \lambda _3 \mu_1
  ^2}{\lambda _1 \lambda _2 - \frac{1}{4} \lambda _3^2} \, .
\label{minima}
\end{equation}
To compute the scalar masses, we must expand the potential
(\ref{higgsV}) around the minima (\ref{minima}). We denote by $h$ and
$h''$ the scalar fields of definite masses, $m_{h}$ and $m_{h''}$
respectively. After a bit of algebra, the explicit
expressions for the scalar mass eigenvalues and eigenvectors are given
by
\begin{eqnarray}
m^2_{h} &=& \lambda _1 v_1^2 + \lambda _2 v_2^2 - \sqrt{(\lambda _1
  v_1^2 - \lambda _2 v_2^2)^2 + (\lambda _3 v_1 v_2)^2} \, , \label{mh1}\\
m^2_{h''} &=& \lambda _1 v_1^2 + \lambda _2 v_2^2 + \sqrt{(\lambda _1 v_1^2 - \lambda _2 v_2^2)^2 + (\lambda _3 v_1v_2)^2} \, ,
 \label{mh2}
\end{eqnarray}
\begin{equation}
\left( \begin{array}{c} h\\h''\end{array}\right) = \left( \begin{array}{cc} \cos{\alpha}&-\sin{\alpha}\\ \sin{\alpha}&\cos{\alpha}
	\end{array}\right) \left( \begin{array}{c} h_1\\h_2\end{array}\right) \, ,
\end{equation}
where $\alpha \in [-\pi/2, \pi/2]$ also  fullfils
\begin{eqnarray}\label{sin2a}
\sin{2\alpha} &=& \frac{\lambda _3 v_1v_2}{\sqrt{(\lambda _1 v_1^2 - \lambda _2 v_2^2)^2 + (\lambda _3 v_1v_2)^2}} \, ,\\ 
\cos{2\alpha} &=& \frac{\lambda _1 v_1^2 - \lambda _2 v_2^2}{\sqrt{(\lambda _1 v_1^2 - \lambda _2 v_2^2)^2 + (\lambda _3 v_1v_2)^2}}\, .
\end{eqnarray}
Now, it is convenient to invert (\ref{mh1}), (\ref{mh2}) and (\ref{sin2a}), 
to extract the parameters in the Lagrangian in terms of the physical quantities $m_{h}$, $m_{h''}$ and $\sin{2\alpha}$
\begin{eqnarray}
\label{12}
\lambda _1 &=& \frac{m_{h''}^2}{4v_1^2}(1-\cos{2\alpha}) +
\frac{m_{h}^2}{4v_1^2}(1+\cos{2\alpha}), \nonumber \\ 
\lambda _2 &=& \frac{m_{h}^2}{4v_2^2}(1-\cos{2\alpha}) +
\frac{m_{h''}^2}{4 v_2^2}(1+\cos{2\alpha}),\\ 
\lambda _3 &=& \sin{2\alpha} \left( \frac{m_{h''}^2-m_{h}^2}{2v_1v_2}
\right). \nonumber
\end{eqnarray}

One-loop corrections to (\ref{higgsV}) can be implemented by
  making $\lambda_1$, $\lambda_2$, and $\lambda_3$ field dependent
  quantities. Equation (\ref{VernonGabePaul}) then needs to be imposed
  in the regions where this is the case. When we talk about the
  stability of (\ref{higgsV}) at some energy $Q$ (with the use of the
  couplings at that scale), we are thinking that the field values are
  at the scale $Q$. Note that the field values are the only physical
  quantities when talking about a potential like (\ref{higgsV}), and
  therefore the appropriate renormalization scale must also be at that
  scale. For $\lambda_3>0$ the third condition in
  (\ref{VernonGabePaul}) is only invalidated for field values $v_1$
  around $m_{h''}$ regardless of the renormalization scale
  $Q$~\cite{EliasMiro:2012ay}.  Namely, the instability region is
  given by
\begin{equation}
v_2 < \frac{m_{h''}}{\sqrt{2\lambda_2}}, \quad   Q_- < v_1 < Q_+,\quad
    \left. Q^2_\pm = \frac{m_{h''}^2 \lambda_3}{8 \lambda_1 \lambda_2} \left(1
  \pm \sqrt{1 - \frac{4\lambda_1 \lambda_2}{\lambda_3^2}} \right) \right|_{Q_*} \,,
\label{rolfi}
\end{equation}
where $Q_*$ is some energy scale where the extra positivity condition
is violated~\cite{EliasMiro:2012ay}. Thus, $Q_\pm \sim m_{h''}$ when the extra positivity
condition is saturated, i.e. $\lambda_1 \lambda_2 =
\lambda_3^2/4$. From (\ref{rolfi}) we see that $Q_\pm \sim m_{h''}$
when all the $\lambda_i$ are roughly at the same scale. If one of the
$\lambda_{1,2}$ is close to zero, then $Q_+$ can be $\gg m_{h''}$, but
this region of the parameter space is taken care of by the condition
$\lambda_{1,2}>0$. The stability for field values at $m_{h''}$ is then
determined by the potential with coupling at scale $m_{h''}$ (instead
of $Q$). Therefore, for $\lambda_3>0$, we impose the extra positivity
condition in the vicinity of $m_{h''}$.  Even though the potential
appears to be unstable at $Q \gg m_{h''}$, it is actually stable when
all the field values are at the scale $Q$. Note that the potential
with $\lambda_i(Q)$ can only be used when the physical quantities
(field values $v_1$, $v_2$) are at the scale $Q$. On the other hand,
the instability region for $\lambda_3<0$ reads
\begin{equation}
v_2 > \frac{m_{h''}}{\sqrt{2\lambda_2}}, \quad c_- < \frac{v_1}{v_2}
  <c_+, 
    \quad \left. c_{\pm}^2 = - \frac{\lambda_3}{2\lambda_1} \left( 1 \pm \sqrt{1
        - \frac{4\lambda_1 \lambda_2}{\lambda_3^2} }\right) \right|_{Q_*}\,,
\end{equation}
and hence is given by the ratio of $v_1$ and $v_2$, which can be
reached even with both $v_1$ and $v_2$ being $\gg
m_{h''}$~\cite{EliasMiro:2012ay}. Therefore, for $\lambda_3< 0$, we
impose the extra positivity condition at all energy scales.  Note that
the asymmetry in $\lambda_3$ will carry over into an asymmetry in
$\alpha$.

As we noted above, the fields $C_\mu,  X_\mu, B_\mu$ are related
to $Y_\mu, Y_\mu{}', Y_\mu{}''$ by an Euler rotation 
matrix~\cite{Anchordoqui:2011ag},
\begin{equation}
\mathbb{R}=
\left(
\begin{array}{ccc}
 C_\theta C_\psi  & -C_\phi S_\psi + S_\phi S_\theta C_\psi  & S_\phi
S_\psi +  C_\phi S_\theta C_\psi  \\
 C_\theta S_\psi  & C_\phi C_\psi +  S_\phi S_\theta S_\psi  & - S_\phi
C_\psi + C_\phi S_\theta S_\psi  \\
 - S_\theta  & S_\phi C_\theta  & C_\phi C_\theta
\end{array}
\right) \,.
\end{equation}
Hence, the covariant derivative
for the $U(1)$ fields in Eq.~(\ref{covderi2}) can be rewritten in terms of
$Y_\mu$, $Y'_\mu$, and $Y''_\mu$ as follows
\begin{eqnarray}
\CD_\mu & = & \partial_\mu -i Y_\mu \left(-S_\xt g'_1 Q_{I_R} + C_\theta S_\psi  g'_4  Q_{L} +  C_\theta C_\psi g'_3 Q_B \right) \nonumber \\
 & - & i Y'_\mu \left[ C_\theta S_\phi  g'_1 Q_{I_R} +\left( C_\phi C_\psi + S_\theta S_\phi S_\psi \right)  g'_4 Q_{L} +  (C_\psi S_\theta S_\phi - C_\phi S_\psi) g'_3 Q_B \right] \label{linda} \\
& - & i Y''_\mu \left[ C_\theta C_\phi g'_1 Q_{I_R} +  \left(-C_\psi S_\phi + C_\phi S_\theta S_\psi \right)  g'_4  Q_{L} + \left( C_\phi C_\psi S_\theta + S_\phi S_\psi\right) g'_3 Q_B \right]   \, .  \nonumber
\end{eqnarray}
Now, by demanding that $Y_\mu$ has the
hypercharge 
\begin{equation}
Q_Y = c_1 Q_{I_R} + c_3 Q_B + c_4 Q_L   
\label{hyperchargeY}
\end{equation}
 we  fix the first column of the rotation matrix $\mathbb{R}$
\begin{equation}
\bay{c} C_\mu \\  X_\mu \\ B_\mu
\eay = \left(
\begin{array}{lr}
  Y_\mu \,  c_3 g_Y/g'_3& \dots \\
  Y_\mu \,  c_4 g_Y/g'_4 & \dots\\
   Y_\mu \, c_1 g_Y/g'_1 & \dots
\end{array}
\right) \, ,
\end{equation}
and we determine the value of the two associated Euler angles
\begin{equation}
\theta = {\rm -arcsin} [c_1 g_Y/g'_1]
\label{theta}
\end{equation}
and
\begin{equation}
\psi = {\rm arcsin}  [c_4 g_Y/ (g'_4 \, C_\theta)] \, ,
\label{psi}
\end{equation}
with $c_1 = 1/2$, $c_3 =1/6$, $c_4 = -1/2$, $B=Q_B/3$ and  $L=Q_{L}$.
The couplings $g'_1$ and $g'_4$ are related through the orthogonality
condition, $P(g_Y,g'_1,g'_3,g'_{4}) =0$, yielding
\begin{equation}
 \left(\frac{c_4}{ g' _4} \right)^2  = \frac{1}{g_Y^2} - \left(\frac{c_3}{g'_3} \right)^2  - \left(\frac{c_1}{g'_1}\right)^2  \, ,
\label{orthogonality25}
\end{equation}
with $g'_3$ fixed by the relation of $U(N)$ unification
\begin{equation}
g'_3 (M_s) = \frac{1}{\sqrt{6}} \, g_3(M_s) \, .
\label{U(N)}
\end{equation}
 Next, by demanding that $Y''_\mu$
couples to an anomalous free linear combination of $I_R$ and $B-L$ we
determine the third Euler angle
\begin{equation}
\tan \phi = - S_\theta \frac{3 \ g'_3 \ C_\psi + g'_4 \ S_\psi}{3 \ g'_3 \ 
  S_\psi - g'_4 \ C_\psi}  \  .
\label{tanfi}
\end{equation}
The absence of abelian, mixed, and mixed gauge-gravitational anomalies
is ensured utilizing the generalized Green-Schwarz mechanism, in which
triangle anomalies are cancelled by Chern-Simons couplings. In the
$Y$-basis we require the $({\rm mass})^2$ matrix of the anomalous
sector to be ${\rm diag} (0, {M'}^2, 0)$. For the heavy field we take
$M' \sim M_s$ and therefore $Y'_\mu \simeq Z'_\mu$ decouples from the low
energy physics. The non-anomalous gauge boson, $Y''_\mu \simeq Z''_\mu$ grows
a TeV-scale mass via $H''$~\cite{Anchordoqui:2012wt}.

Altogether, the covariant derivative of the low energy effective theory reads
\begin{eqnarray}
 {\cal D}_\mu & = & \partial_\mu - i g_3 T^a G^a_\mu -  i g_2 \tau^a W^a_\mu - i g_Y \ Q_Y \ Y_\mu  -  i g_{\cal Y} \
Q_Y \ Y''_\mu   -  i g_{B-L} \, (B-L) \ Y''_\mu ,
\label{newcovderi}
\end{eqnarray} 
where
\begin{eqnarray}
\label{ic}
g_Y  & = & - \tfrac{1}{2} S_\theta \, g'_1 = - \tfrac{1}{2}  C_\theta
S_\psi g'_4 = \tfrac{3}{2} C_\theta C_\psi
g'_3 \nonumber \\
g_{\cal Y} & = & 2 \, C_\theta \, C_\phi \, g'_1 \\
g_{B-L} & = & 3 g'_3 (C_\phi C_\psi S_\theta + S_\phi S_\psi) -
C_\theta C_\phi g'_1 . \nonumber
\end{eqnarray}
Finally, a straightforward calculation leads to the RG equations for the five
parameters in the scalar potential~\cite{Basso:2010jm}
\begin{eqnarray} \label{RG}
 \frac{d \mu_1^2}{dt} &=&
\frac{\mu_1^2}{16\pi ^2}\left( 12\lambda _1 +6Y_t^2+2\frac{\mu_2
    ^2}{\mu_1^2}\lambda _3
  -\frac{9}{2}g_2^2-\frac{3}{2}g_Y^2-\frac{3}{2} g_{\cal Y}^2\right)\,
, \nonumber \\ 
\frac{d \mu_2 ^2}{dt} &=& \frac{\mu_2^2}{16\pi
   ^2}\left( 8\lambda _2 +4\frac{\mu_1^2}{\mu_2^2}\lambda _3 - 24
   g_{B-L}^{2}\right)\, , \nonumber \\  \frac{d\lambda_1}{dt} & = &
 \frac{1}{16\pi ^2}\left( 24\lambda _1^2+\lambda _3^2
-6Y_t^4 +\frac{9}{8}g_2^4 +\frac{3}{8}g_Y^4 +\frac{3}{4}g_2^2g_Y^2
+\frac{3}{4}g_2^2
g_{\cal Y}^2  +\frac{3}{4}g_Y^2 g_{\cal Y}^2+\frac{3}{8} g_{\cal Y}^4 \right.
\nonumber \\ & + &
\left.  12\lambda _1 Y_t^2 
 -   9\lambda _1 g_2^2-3\lambda _1 g_Y^2-3\lambda _1
  g_{\cal Y}^2
	\right)\, ,\\ 
 \frac{d \lambda _2}{dt} &=&  \frac{1}{8\pi ^2}\left( 10\lambda
   _2^2+\lambda _3^2 +48 g_{B-L}^{4} -24\lambda _2g_{B-L}^{2}
 \right)\, , \nonumber \\ 
\frac{d \lambda _3}{dt} &=&  \frac{\lambda _3}{8\pi
  ^2}\left( 6\lambda _1+4\lambda _2+2\lambda
  _3+3Y_t^2-\frac{9}{4}g_2^2-\frac{3}{4}g_Y^2-\frac{3}{4} g_{\cal Y}^2
  - 12 g_{B-L}^{2} \right) + \frac{3}{4 \pi^2} \, g_{\cal Y}^2 \, g_{B-L}^{2}  \nonumber ,
\end{eqnarray}
where $t = \ln Q$ and $Y_t$ is the top Yukawa coupling, with
\begin{equation}\label{RGE_yuk_top}
\frac{dY_t}{dt} = \frac{Y_t}{16\pi ^2}\left(
  \frac{9}{2}Y_t^2-8g_3^2-\frac{9}{4}g_2^2-\frac{17}{12}g_Y^2-\frac{17}{12}
  g_{\cal Y}^2 -\frac{2}{3}g_{B-L}^{2}-\frac{5}{3} g_{\cal Y} g_{B-L} \right)
\end{equation}
and $Y_t^{(0)} = \sqrt{2} \, m_t/v$.
The RG running of the gauge couplings follow the standard form
\begin{eqnarray}
\frac{dg_3}{dt} &=& \frac{g_3^3}{16\pi ^2}\left[ -11+\frac{4}{3}n_g
\right] = - \frac{7}{16} \, \frac{ g_3^3}{\pi ^2} \, , \nonumber \\
\frac{dg_2}{dt} &=& \frac{g_2^3}{16\pi ^2}\left[
  -\frac{22}{3}+\frac{4}{3}n_g+\frac{1}{6}\right] = -\frac{19}{96}
\,\frac{g_2^3}{\pi ^2} \,,  \nonumber \\
\frac{dg_Y}{dt}  &=& \frac{1}{16\pi ^2}\left[A^{YY}g_Y^3 \right]\, , \\ 
\frac{dg_{B-L}}{dt} &=& \frac{1}{16\pi ^2}\left[A^{(B-L)
    (B-L)}g_{B-L}^3+2A^{(B-L) Y}g_{B-L}^2 g_{\cal Y}+A^{YY}g_{B-L}
  g_{\cal Y}^2 \right] \, , \nonumber \\ 
\frac{dg_{\cal Y}}{dt} &=& \frac{1}{16\pi ^2}\left[A^{YY} g_{\cal
    Y}\,(g_{\cal Y}^2+2g_{Y}^2)+2A^{(B-L)Y}g_{B-L}(g_{\cal
    Y}^2+g_{Y}^2)+A^{(B-L) (B-L)}g_{B-L}^2 g_{\cal Y} \right] ,
\nonumber 
\end{eqnarray}
where $n_g =3$ is the number of generations and
\begin{equation}\label{A_charge}
A^{ab} = A^{ba} = \frac{2}{3} \sum _f Q_{a,f} Q_{b,f} + \frac{1}{3}\sum _s
Q_{a,s} Q_{b,s}\, , \qquad (a,b=Y,\ B-L) \,,
\end{equation}
with $f$ and $s$ indicating
contribution from fermion and scalar loops, respectively.

For energies below the mass of the heavier Higgs $H''$, the
  effective theory is (of course) the SM. In the low energy regime the
  scalar Lagrangian then reads
\begin{equation}
\mathscr{L}_s =\left( {\cal D}^{\mu} H\right) ^{\dagger} {\cal
  D}_{\mu} H - \mu^2 \left| H \right|^2 - \lambda \left| H \right|^4 \,,
\label{higgsSM}
\end{equation}
and the RG equations are those of SM. To obtain the matching
conditions connecting the two theories,
following~\cite{EliasMiro:2012ay} we integrate out the field $H''$ to
obtain a Lagrangian of the form (\ref{higgsSM}). Identifying the quadratic
and quartic terms in the potential yields 
\begin{equation}
\mu^2 = \mu_1^2 - \mu_2^2 \ \frac{\lambda_3}{2 \lambda_2}  
\end{equation}
and
\begin{equation}
\lambda = \lambda_1 \ \left(1 - \frac{\lambda_3^2}{4 \lambda_1 \lambda_2} \right) \, ,
\end{equation}
respectively. This is consistent with the continuity of $v \leftrightharpoons 
v_1$; namely
\begin{equation}
v^2 = - \left. \frac{\mu^2}{\lambda} \right|_{Q = m_{h''}} = -
\left. \frac{\mu_1^2 - \mu_2^2 \ \lambda_3 /(2 \lambda_2)}{ \lambda_1
    \ 
\left[1 - \lambda_3^2/(4 \lambda_1 \lambda_2) \right]} \right|_{Q = m_{h''}} \, ,
\end{equation}
or equivalently
\begin{equation}
\left. v^2 \right|_{Q = m_{h''}} = \left. v_1^2 \right|_{Q = m_{h''}} \,,
\end{equation}
with $v_1$ given by (\ref{minima}). The quartic interaction between
the heavy scalar singlet and the Higgs doublet provides an essential
contribution for the stabilization the scalar field
potential~\cite{EliasMiro:2012ay}.

\section{Results and Conclusions}
\label{III}

To ensure perturbativity of $g'_4$ between the TeV scale and the
string scale we find from (\ref{orthogonality25}) that $g'_1 >
0.232$. We also take $g'_1 \alt 1$ in order to ensure perturbativity
at the string scale. Let us first study the region of the parameter
space constrained by $g'_1 (M_s) \simeq 1$. The string-scale values
of the other abelian couplings are fixed by previous considerations
(\ref{orthogonality25}) and (\ref{U(N)}). The Euler angles at $M_s$
are also fixed by (\ref{theta}), (\ref{psi}), and (\ref{tanfi}).  All
the couplings and angles are therefore determined at all energies
through RG running. As an illustration we set $M_s = 10^{14}~{\rm
  GeV}$; this leads to $g'_3 (M_s)= 0.231$, $g'_4 (M_s) = 0.232$,
$\psi (M_s) = -1.245$, $\theta (M_s) = -0.217$, and $\phi (M_s) =
-0.0006$. Next, we define $Q_{\rm min} = 125~{\rm GeV}$ and normalize $t
= \ln(Q/125~{\rm GeV})$ and $t_{\rm max} = \ln(\Lambda/125~{\rm
  GeV})$. Finally, we run the couplings and angles down to the TeV
region: $g'_1 = 0.406$, $g'_3 = 0.196$, $g'_4 = 0.218$, $\theta =
-0.466$, $\psi = -1.215$, and $\phi = -0.0003$.

Using the SM relation $m_H^2 = - 2 \mu^2$, with $m_H \simeq 125~{\rm
  GeV}$, and setting $v^2 = 246~{\rm GeV}$ at the same energy scale $Q
= 125~{\rm GeV}$ fixes the initial conditions for the parameters $\mu$
and $\lambda$. It should be noted that we take the top Yukawa coupling
evaluated at $m_t$. This introduces a small unnoticeable error.  On
the other hand, $m_t$ is taken to be the physical top mass. Have we
used the running mass instead as in~\cite{Casas:1994us}, the running
of the quartic coupling $\lambda$ would become much slower, with the
instability scale pushed to almost $10^9~{\rm GeV}$.  We run the SM
couplings from 125~GeV up to the mass scale $m_{h''}$ and use the
matching conditions to determine $v$, which in turns allows one to
solve algebraically for $m_h$.

\begin{figure}[tbp] 
\begin{minipage}[t]{0.49\textwidth}
    \postscript{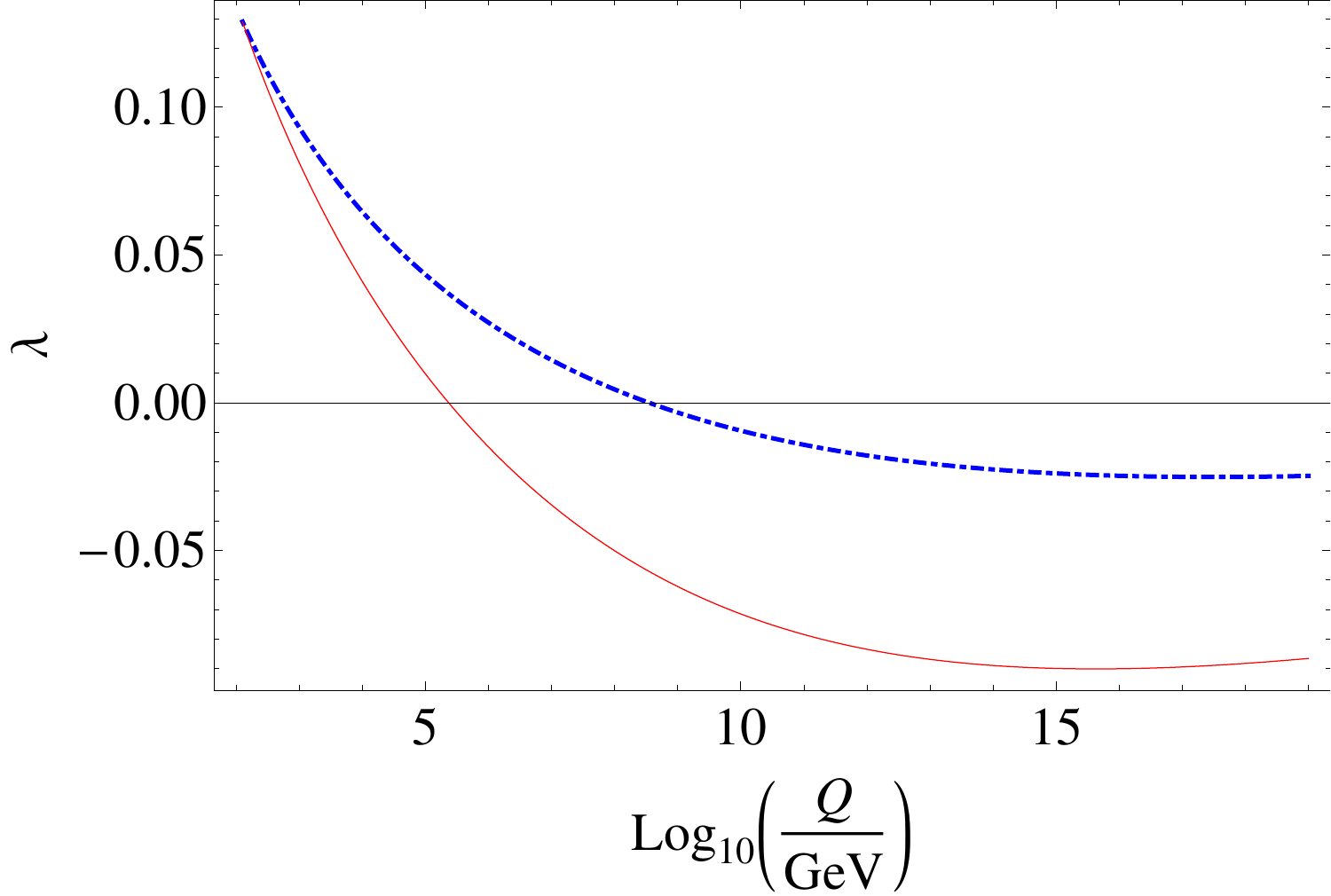}{1.02} \end{minipage}
  \begin{minipage}[t]{0.49\textwidth}
    \postscript{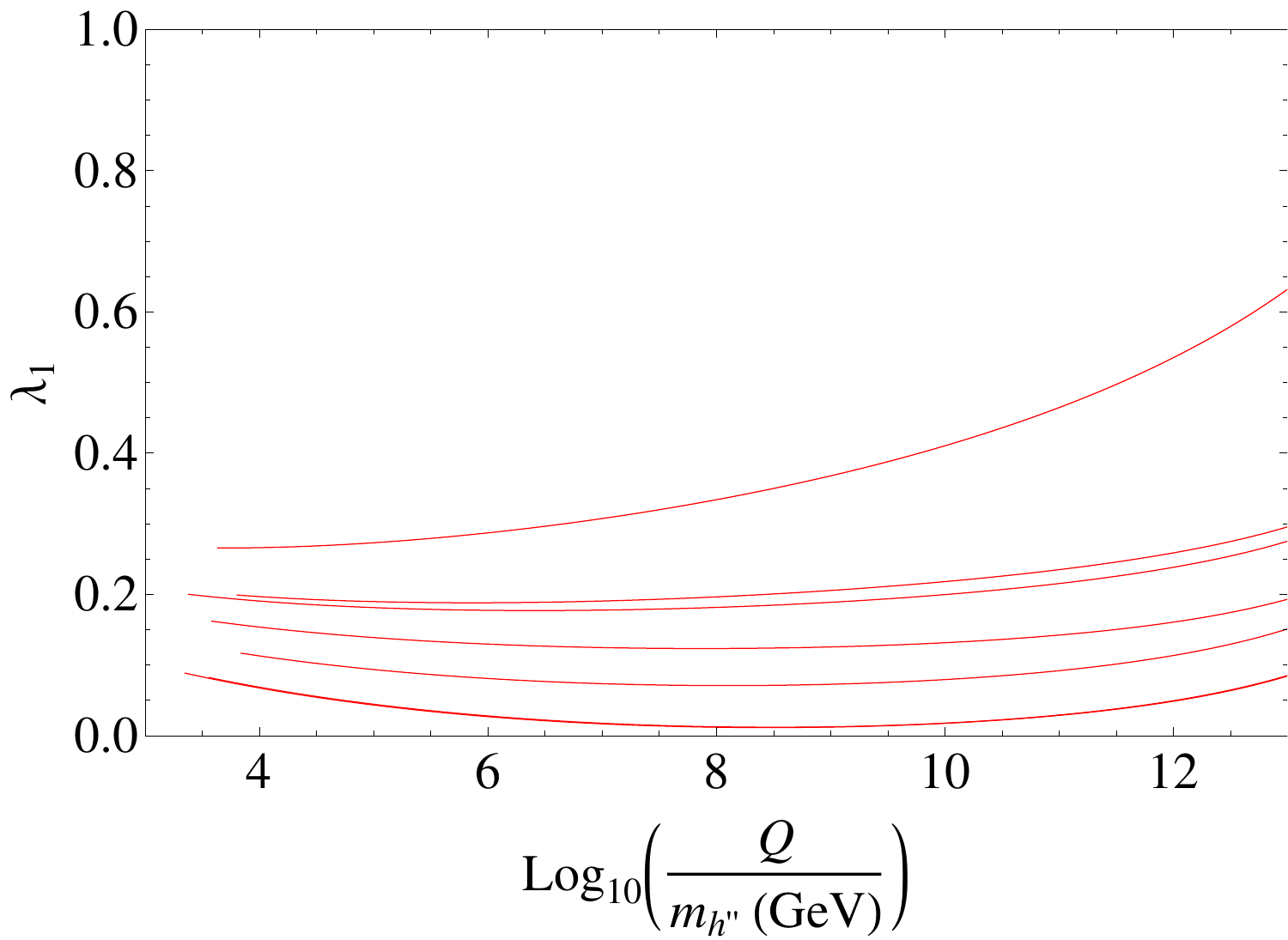}{0.97} \end{minipage} 
\begin{minipage}[t]{0.49\textwidth}
    \postscript{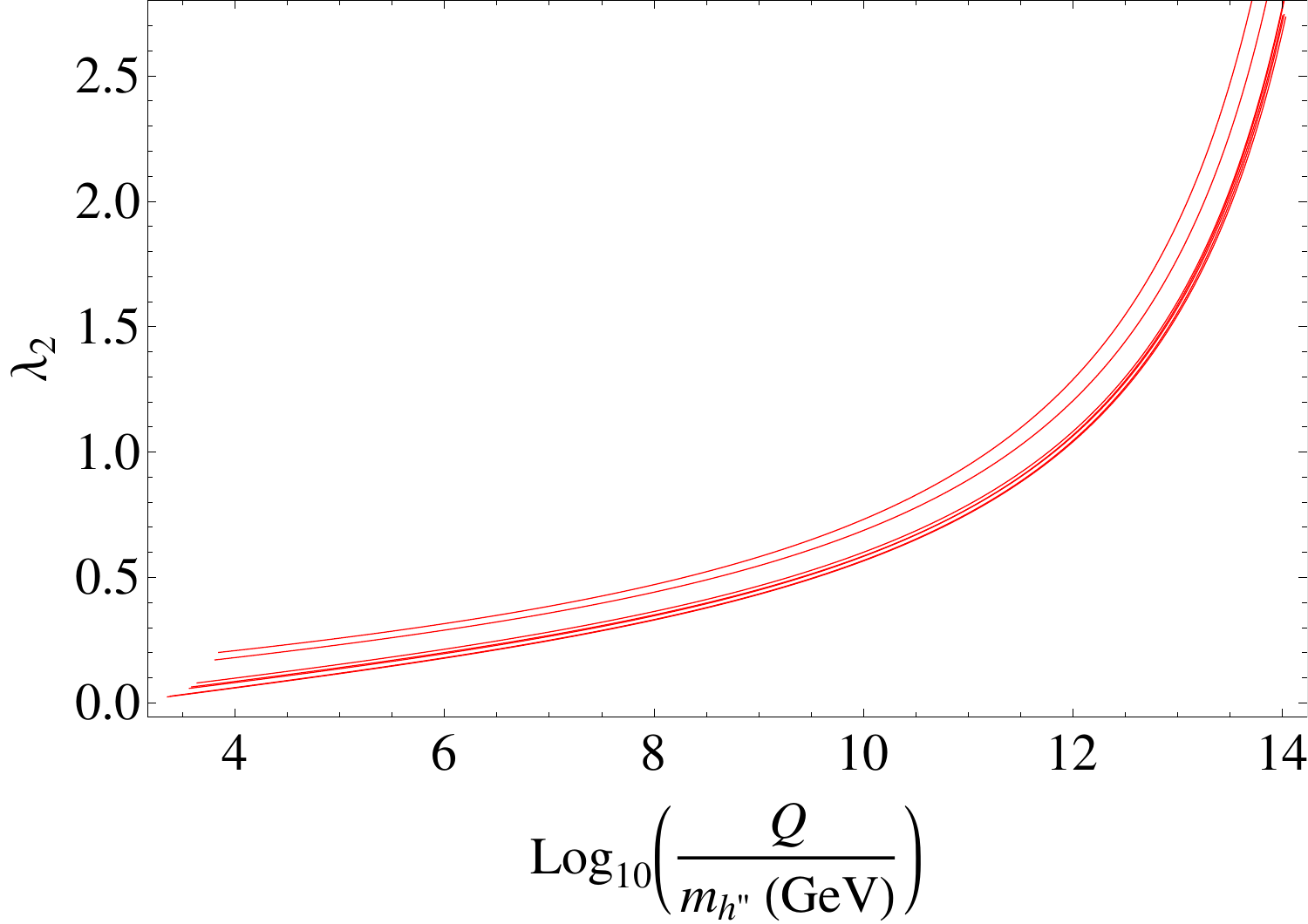}{0.97} \end{minipage} 
  \begin{minipage}[t]{0.49\textwidth}
    \postscript{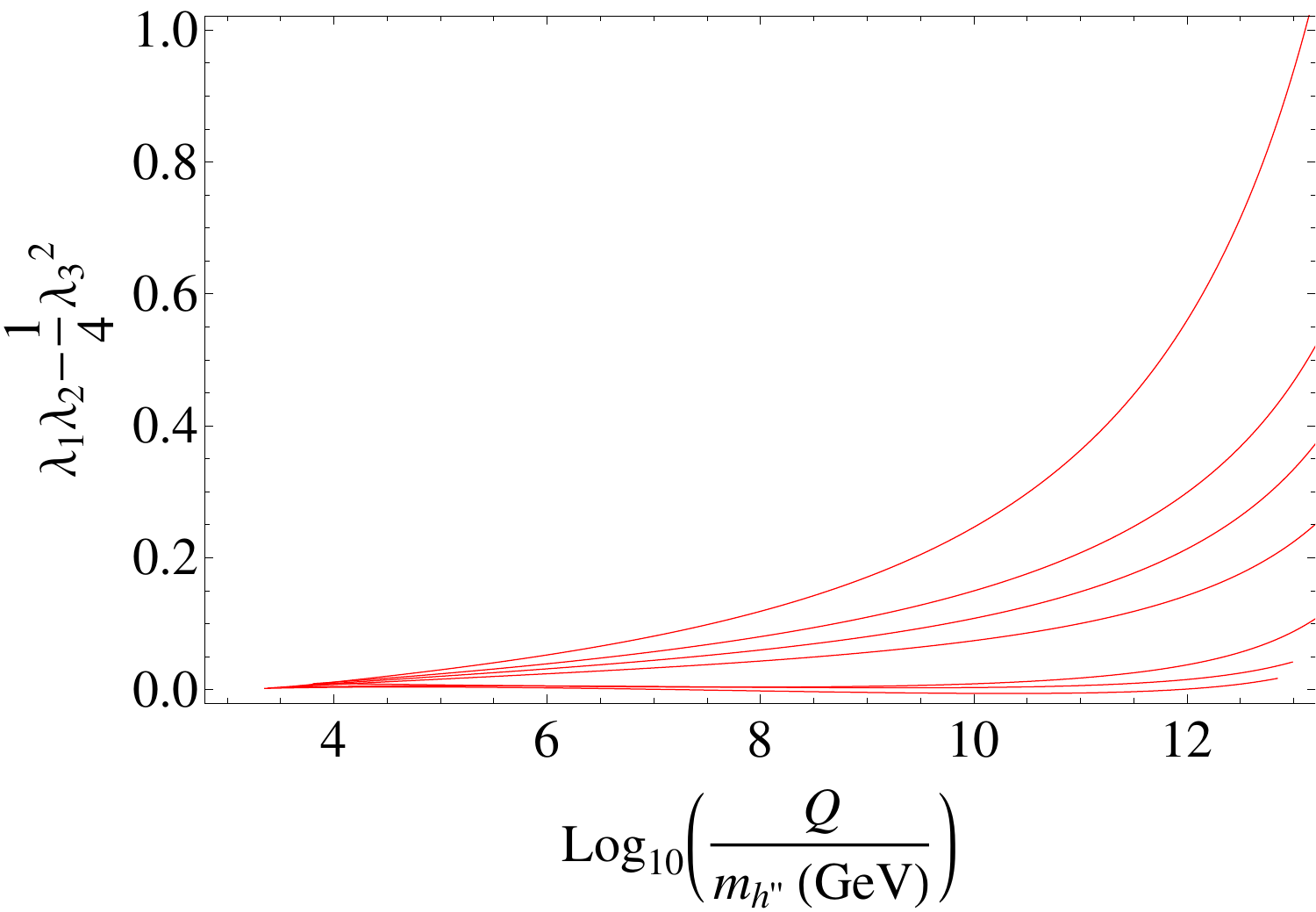}{1.02} \end{minipage} 
  \caption{From left to right downwards: the first panel shows the
    running of $\lambda$ from its value at $125~{\rm GeV}$ (red solid
    line $m_t = 172.9~{\rm GeV}$ and blue dot-dashed line $m_t =
    164~{\rm GeV}$); the second and third panels show the typical
    behavior of the running couplings $\lambda_1(t)$ and $\lambda_2
    (t)$ for the average value of the initial condition, $\langle
    \lambda_1^{(0)} \rangle = 0.28$ in the integration of (\ref{RG});
    the fourth panel shows the behavior of the extra positivity
    condition for $\alpha <0$. In the running of $\lambda_i$ we have
    taken $M_s = 10^{14}~{\rm GeV}$.}
\label{sm++_f3} 
\end{figure}

\begin{figure}[tbp] 
    \postscript{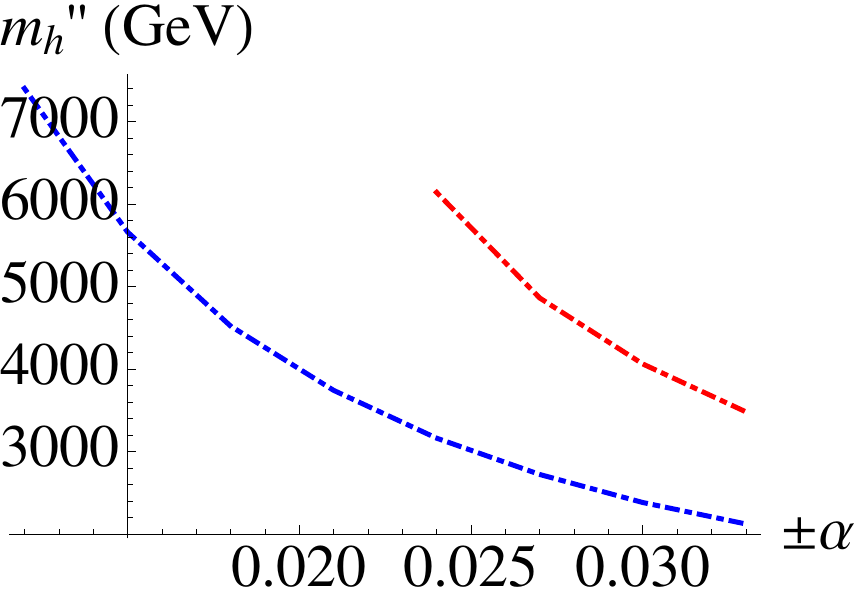}{0.8} 
    \caption{The lower boundary of the allowed parameter space in the
      $m_{h''}-\alpha$ plane under the vacuum stability constraint of
      Eq.~(\ref{VernonGabePaul}), for the positive alpha (blue) and negative alpha
      (red). We have taken $M_{Z''} = 4.5~{\rm TeV}$ and $M_s = 10^{14}~{\rm GeV}$.}
\label{sm++_f4} \end{figure}

\begin{figure}[tbp] 
    \postscript{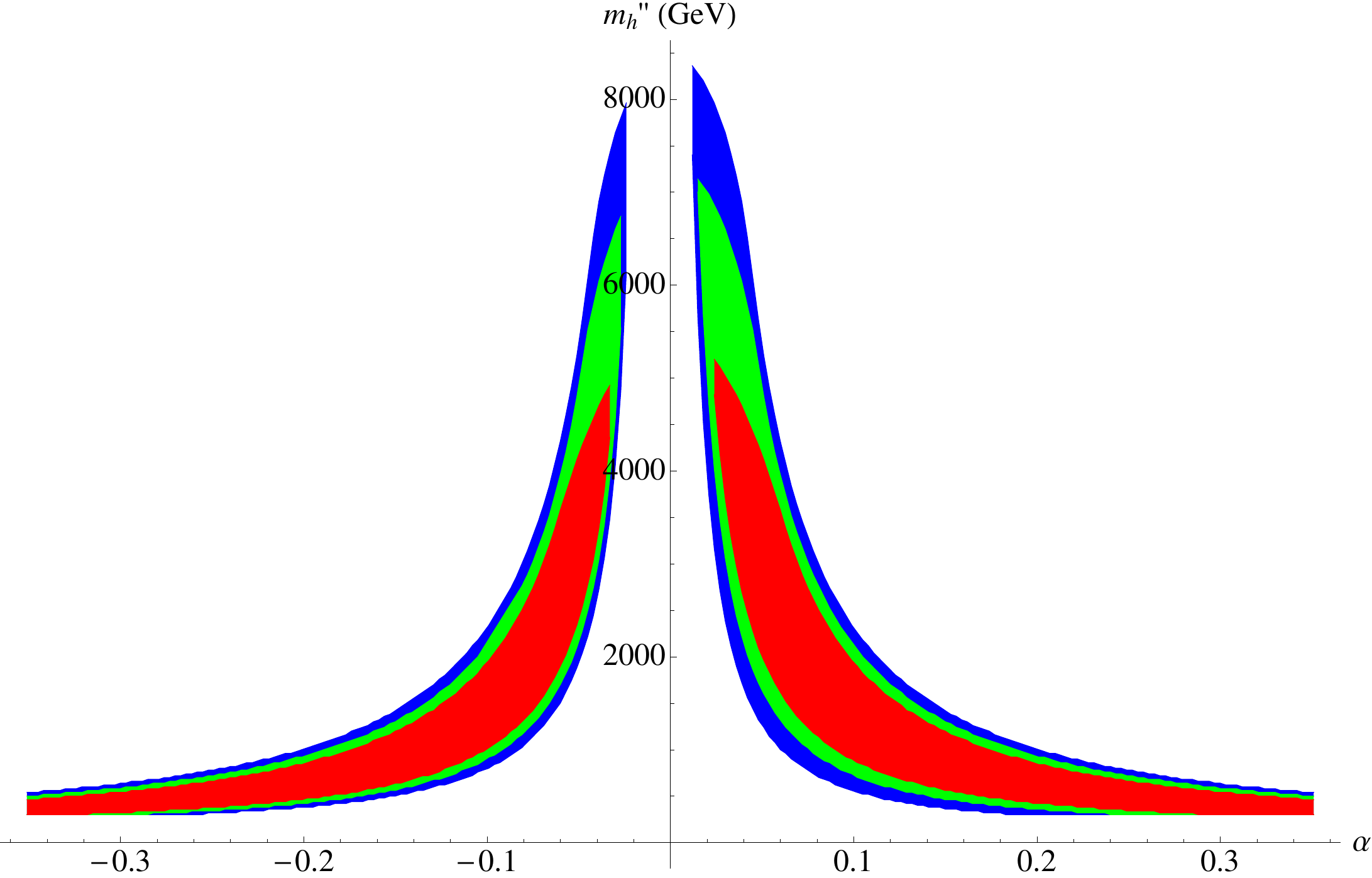}{0.8}
    \caption{The allowed SM$^{++}$ parameter space in the $m_{h''}$ vs
      $\alpha$ plane under the vacuum stability constraint of
      Eq.~(\ref{VernonGabePaul}), for the case $M_{Z''} = 4.5~{\rm
        TeV}$, with $M_s = 10^{14}~{\rm GeV}$ (blue), $M_s =
      10^{16}~{\rm GeV}$ (green), and $M_s = 10^{19}~{\rm GeV}$
      (red). The perturbative upper bound is defined by $\lambda_i <
      2 \pi$.}
\label{sm++_f5} 
\end{figure}

\begin{figure}[tbp] 
    \postscript{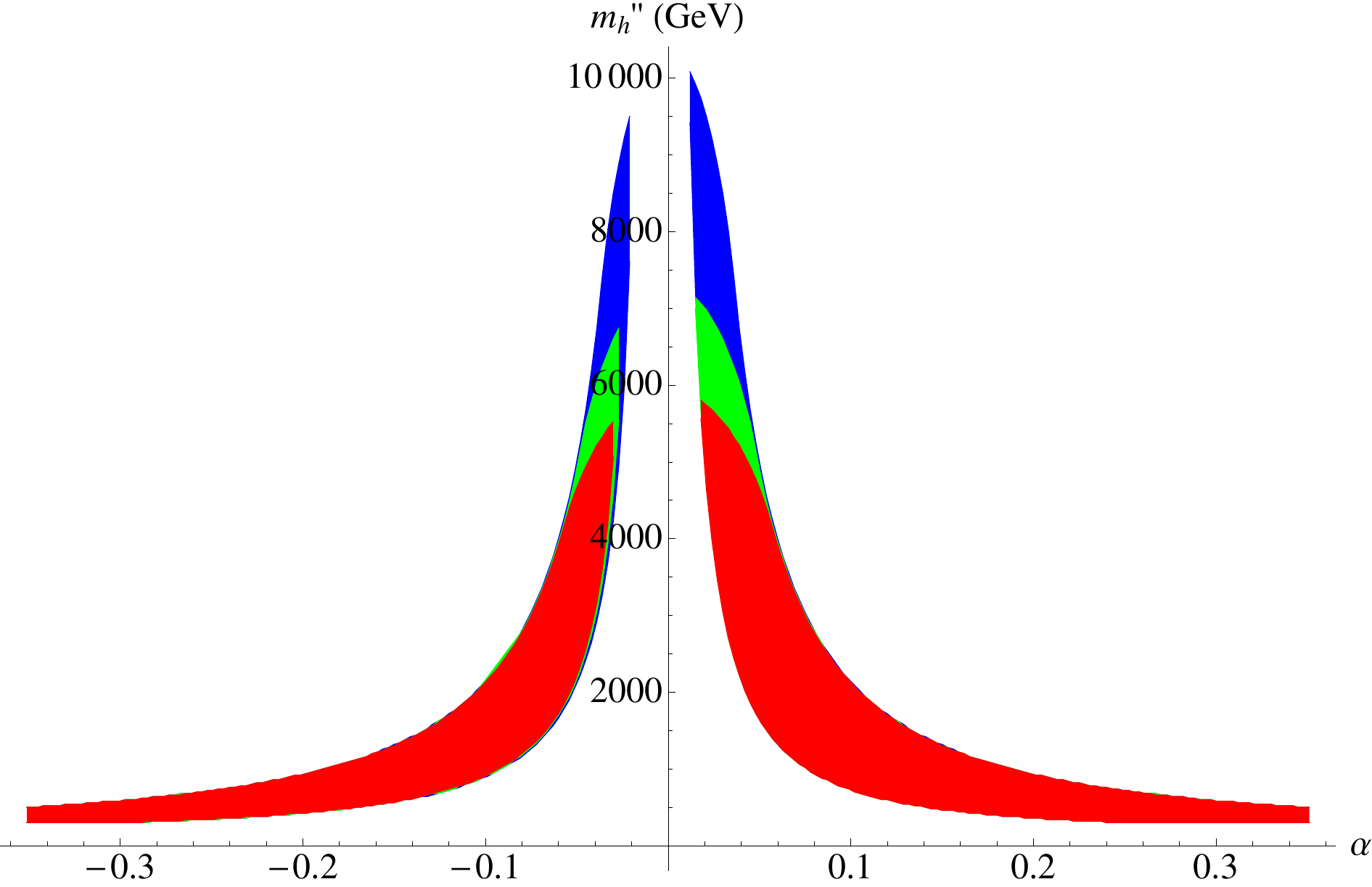}{0.8}
    \caption{Variation of SM$^{++}$ vacuum stability regions with
      $M_{Z''}$. We have taken $M_s = 10^{16}~{\rm GeV}$, $M_{Z''} =
      3.5~{\rm TeV}$ (red), $M_{Z''} = 4.5~{\rm TeV}$ (green), and
      $M_{Z''} = 6.0~{\rm TeV}$ (blue). The perturbative upper bound
      is defined by $\lambda_i < 2 \pi$.}
  \label{sm++_f6}
\end{figure}

\begin{figure}[tbp] 
    \postscript{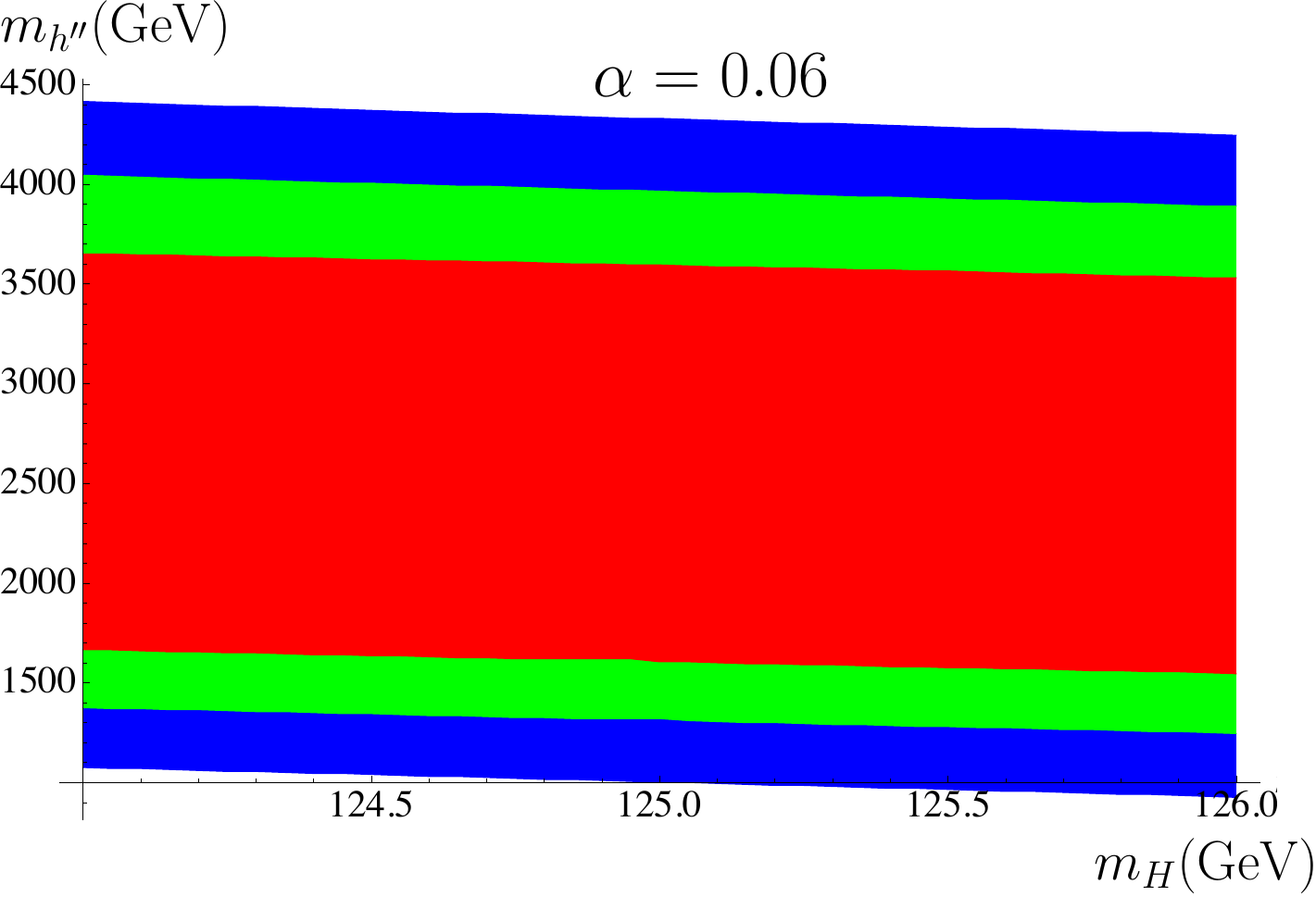}{0.8} 
\caption{Variation of SM$^{++}$ vacuum stability regions with
    $m_H$. We have taken $\alpha = 0.06$, $M_{Z''} = 4.5~{\rm TeV}$,
    $M_s = 10^{14}~{\rm GeV}$ (blue), $M_s = 10^{16}~{\rm GeV}$
    (green), $M_s = 10^{19}~{\rm GeV}$ (red).  The perturbative upper bound is defined by $\lambda_i < 2 \pi$.}
  \label{sm++_f7}
\end{figure}

\begin{figure}[tbp] 
    \postscript{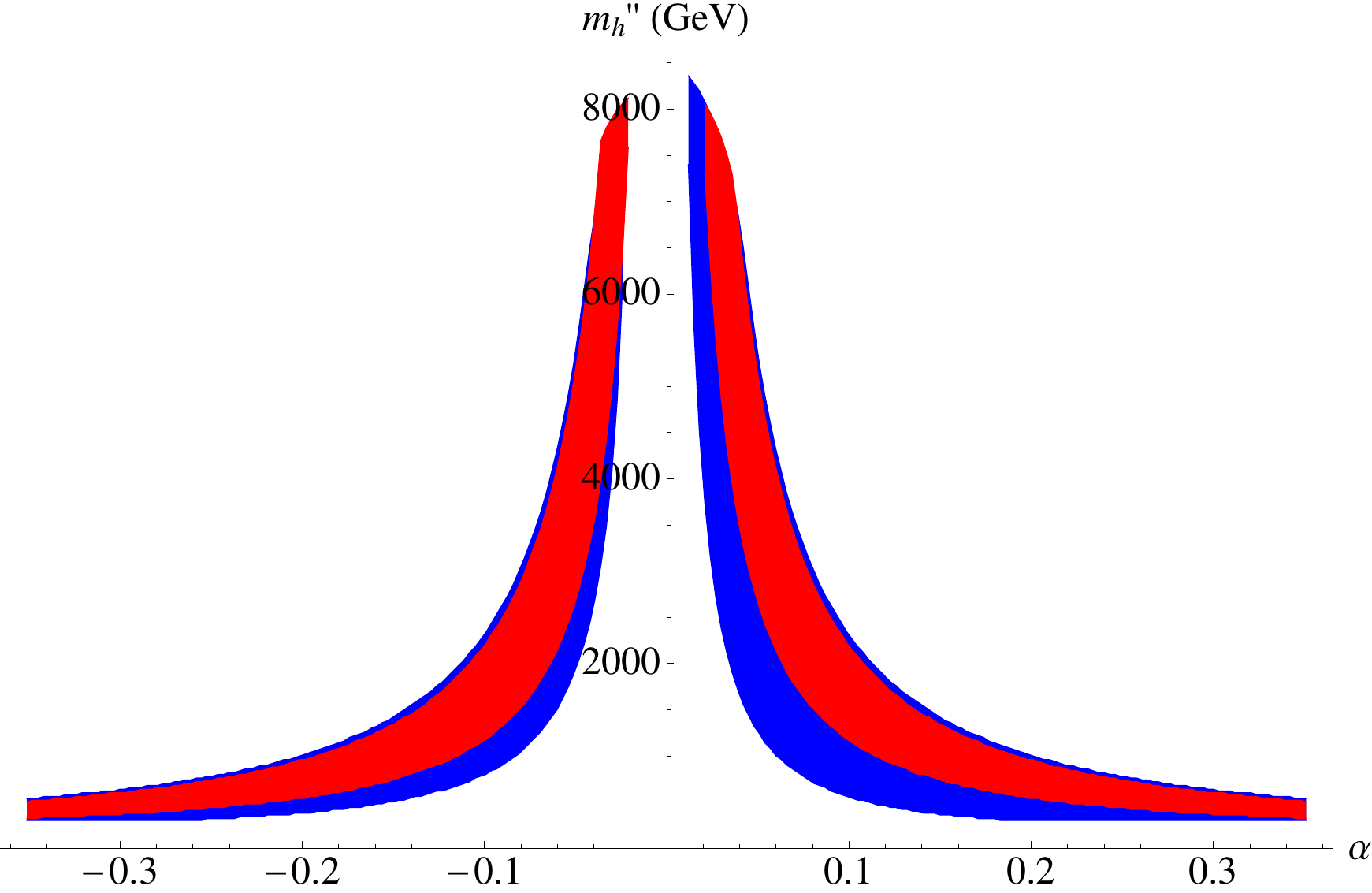}{0.8} 
    \caption{Variation of SM$^{++}$ vacuum stability regions with
      $g'_1(M_s)$. The stable regions correspond to $g_1'(M_s) = 1.000$
      (blue), $g_1'(M_s) =  0.232$ (red). We have taken $M_s =
      10^{14}~{\rm GeV}$, $M_{Z''} = 4.5~{\rm TeV}$, $m_H = 125~{\rm
        GeV}$. The perturbative upper bound is defined by $\lambda_i
      < 2 \pi$.}
  \label{sm++_f8}
\end{figure}

After completing this task, there are {\it a priori} three free
parameters to be fixed at the TeV-scale: $(v_2, \, \alpha,\,
m_{h''})$.  The initial values of $g_Y$, $g_{\cal Y}$ and $g_{B-L}$
are then fixed by previous considerations (\ref{ic}). Actually, using
the relation $M_{Z''} = g'_1 \, C_\phi \,
v_2/C_\theta$~\cite{Anchordoqui:2012wt}, we adopt $(M_{Z''},\,
\alpha,\, m_{h''})$ as the free parameters of the model.\footnote{For
  $M_s=10^{14}~{\rm GeV}$, the $v_2 \leftrightharpoons M_{Z''}$
  relation implies that if $7~{\rm TeV} < v_2 < 13~{\rm TeV}$, then
  $3.2~{\rm TeV} < M_{Z''} < 6.0~{\rm TeV}$. For a different $M_s$
  the range of $M_{Z''}$ is altered because of changes in $g'_1$,
  $\theta$, and $\phi$; {\it e.g.}  for $M_s = 10^{19}~{\rm GeV}$, the
  range becomes $2.8~{\rm TeV} < M_{Z''} < 5.8~{\rm TeV}$.} For $M_s
= 10^{14}$~GeV, we perform a scan of $10^4$ trial random set of
points, $(M_{Z''},\, \alpha,\, m_{h''})$, and using (\ref{12}) we
obtain the initial conditions ($\lambda_1^{(0)}$, $\lambda_2^{(0)}$,
$\lambda_3^{(0)}$) to integrate (\ref{RG}). For each set of points, we
verify that the positivity condition (\ref{VernonGabePaul}) is
fulfilled all the way to $\Lambda = M_s$. The $10^4$ trials are
duplicated for $M_s= 10^{16}$ and $M_s = 10^{19}$~GeV.  Our results
are encapsulated in Figs.~\ref{sm++_f2} to \ref{sm++_f8}, and along
with other aspects of this work are summarized in these concluding
remarks:

\begin{figure}[tbp] 
    \postscript{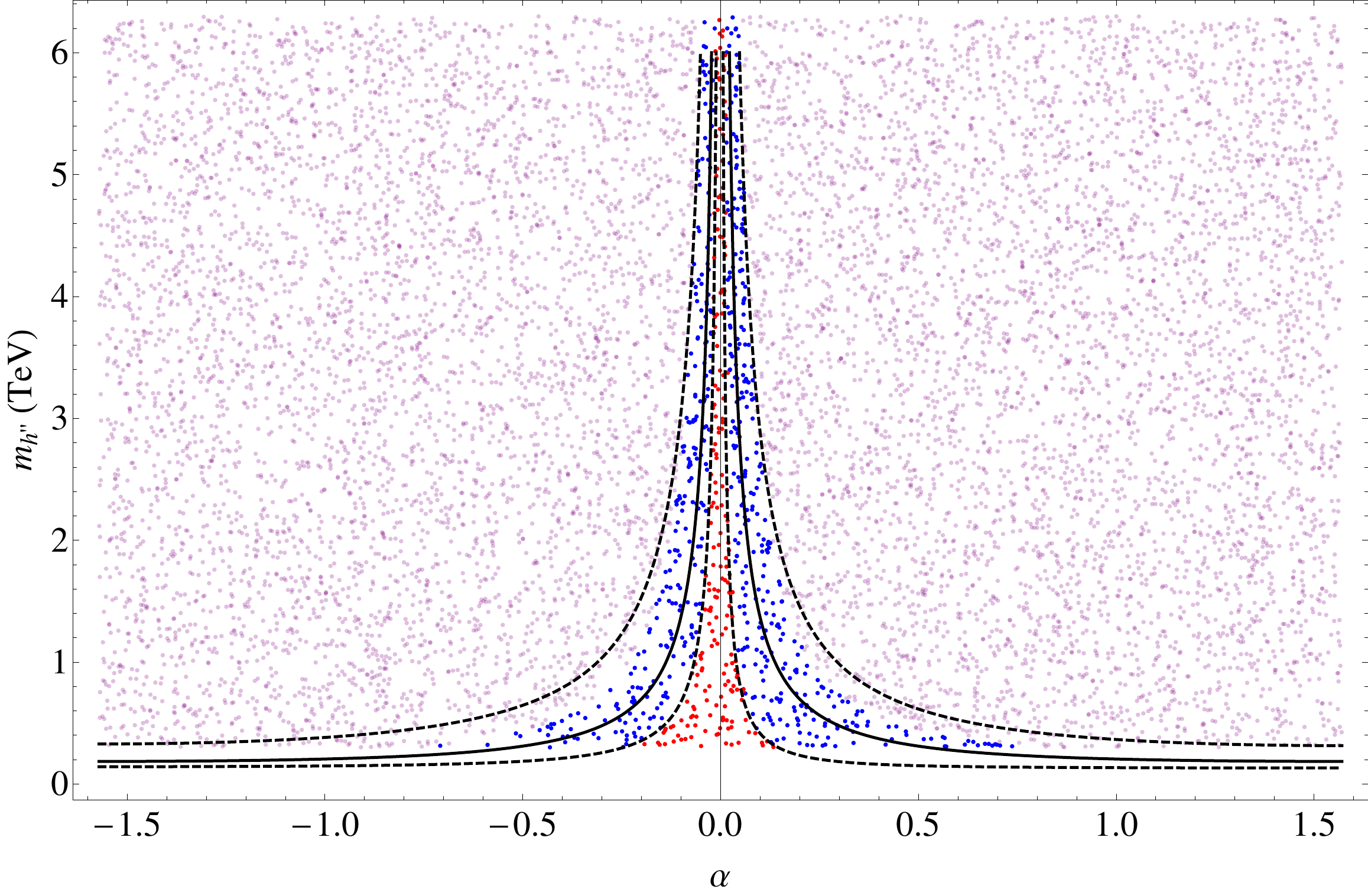}{0.8} 
  \caption{An exhibition of the SM$^{++}$ vacuum stability patterns in
    the $m_{h''}$ vs $\alpha$ plane, for $M_{Z''} = 4.5~{\rm
      TeV}$. The analysis is based on a scan of $10^4$ trial random
    points with $M_s = 10^{14}~{\rm GeV}$. The points yielding a
    stable vacuum solution up to $M_s$ are blue-printed, those leading
    to unstable vacuum solutions are red-printed, and points giving
    runaway solutions ({\em i.e.}, those in which the Higgs doublet
    self-coupling blows up) are purple-printed.
Fits to the  boundaries defining the region with stable
    vacuum solutions (dashed lines) and to the average value of the scatter
    points contained in that region (solid lines) are also shown.}
\label{sm++_f2} \end{figure}

\begin{itemize}
\item In Fig.~\ref{sm++_f2} we show the entire scan for $M_s =
  10^{14}~{\rm GeV}$ and $M_{Z''} = 4.5~{\rm TeV}$. The points yielding a
  stable vacuum solution up to $M_s$ are blue-printed, those leading
  to unstable vacuum solutions are red-printed, and points giving
  runaway solutions are purple-printed.  A stable vacuum solution is
  one in which the positivity condition (\ref{VernonGabePaul}) is
  fulfilled all the way to $\Lambda = M_s$. An unstable solution is
  one in which the stability conditions of the vacuum ($\lambda_1 >
  0$, $\lambda_2 > 0$, $\lambda_1 \lambda_2 > \lambda_3^2/4$) are
  violated.  (Recall that for the case $\lambda_3 >0$ there is no need
  to impose the third condition in (\ref{VernonGabePaul}) at all
  scales, but only in the vicinity of $m_{h''}$.) A runaway solution
  is one in which the RG equations drive the Higgs doublet
  self-coupling non-perturbative. The perturbative upper bound
  (sometimes referred to as `triviality' bound) is given by $\lambda_1
  < 2 \pi$ at any point in the RG evolution~\cite{Ellis:2009tp}.  The
  vacuum stability condition is driven by the behavior of $\lambda_1$,
  and actually is largely dominated by the initial condition
  $\lambda_1^{(0)}$. Indeed, if the extra gauge boson $Z''$ gets its
  mass through a non-Higgs mechanism and the scalar potential
  (\ref{higgsV}) is that of SM ({\em i.e.}  $v_2= \lambda_2 =
  \lambda_3 = 0$), the RG evolution collapses to that of SM and there
  are no stable solutions.\footnote{Of course, even if $v_2 =
    \lambda_2 = \lambda_3 = 0$, with an extra gauge boson the RG
    evolution of $\lambda_1$ is not exactly that of SM, see
    (\ref{RG}).}

\item To determine the range of initial conditions $\lambda_1^{(0)}$
  yielding stable vacuum solutions we fit the boundaries of the blue
  band in the scatter plot. The resulting curves, which are shown as
  dashed lines in Fig.~\ref{sm++_f2}, correspond to $0.16 <
  \lambda_1^{(0)} < 0.96$ for $\alpha <0$ and $0.15 < \lambda_1^{(0)}
  < 0.96$ for $\alpha >0$. The lower limit of $\lambda_1^{(0)}$, which
  defines the boundary between stable and unstable solutions, is close
  to the value required for vacuum stability of the SM potential, as
  shown in (\ref{1}). Namely, substituting $m_h = 130~{\rm GeV}$ and
  $\alpha = 0$ in (\ref{12}) we obtain $\lambda_1^{(0)} = 0.14$.  The
  similarities between the minimum value of $m_H$ that allows absolute
  stability up to the Planck scale within SM and the minimum value of
  $m_h$ in the decoupling limit of (\ref{12}) reinforces our previous
  statement concerning the strong dependence of the RG evolution with
  the initial condition $\lambda_1^{(0)}$.  We have also determined
  the average value of the initial condition $\lambda_1^{(0)}$ through
  a fit to the blue points in the scattered plot. The result, which is
  shown as a solid lines in Fig.~\ref{sm++_f2}, corresponds to
  $\langle \lambda_1^{(0)} \rangle = 0.28$. The behavior of $\lambda$
  together with the typical behavior of $\lambda_1$ and $\lambda_2$
  for the average value of the initial condition $\langle
  \lambda_1^{(0)} \rangle$, are shown in Fig.~\ref{sm++_f3}. Note that
  $\lambda_1$ heads towards the instability and reaches a minimum
  greater than zero; thereafter rises towards the Landau point.  This
  behavior is characteristic of models with scalar
  singlets~\cite{Kadastik:2011aa}. We also show in Fig.~\ref{sm++_f3}
  the typical behavior of $\lambda_1 \lambda_2 -\lambda_3^2/4$ for
  $\alpha <0$ and $\langle \lambda_1^{(0)} \rangle = 0.28$.

\item  Even though the asymmetry between $\pm \alpha$ appears to be
    small on Fig.~\ref{sm++_f2}, it is not actually insignificant. In
    fact, for a given $\alpha$, the lower boundary sometimes changes
    by a factor of two. For example, at $\alpha = 0.24$, it changes from
    $6,140~{\rm GeV}$ to $3,160~{\rm GeV}$. However, the effect on the area is less
    noticeable. The reason is that we can only change the lower
    boundaries of the accepted parameter space. The upper boundary is
    determined by the constraint that $\lambda_i$ (usually $\lambda_2$)
    remains perturbative. This constraint is symmetric with respect to
    $\alpha$. So the area cannot be enlarged indefinitely. Even if
    somehow we can send the lower boundary to zero, the area would only
    increase by another 20\%  to 30\%. The asymmetry becomes more obvious
    if we only consider the lower boundary, as shown in Fig.~\ref{sm++_f4}.

\item To determine the sensitivity of the RG evolution with respect to
  the choice of the string scale, we duplicate the analysis for $M_s =
  10^{16}~{\rm GeV}$ and $M_s = 10^{19}~{\rm GeV}$. The contours display
  in Fig.~\ref{sm++_f5} (for $M_{Z''} = 4.5~{\rm TeV}$) show that the
  region of stable vacuum solutions shrinks as $M_s$ increases.  The
  allowed range of initial conditions with stable vacuum solutions
  therefore depends on the value of the string scale; {\it e.g.}  for
  $M_s = 10^{16}~{\rm GeV}$, we obtain $0.17 < \lambda_1^{(0)} < 0.83$
  for $\alpha < 0$ and $0.16 < \lambda_1^{(0)} < 0.83$ for $\alpha
  >0$, whereas for $M_s = 10^{19}~{\rm GeV}$, we obtain $0.18 <
  \lambda_1^{(0)} < 0.69$ for $\alpha<0$ and $0.17 < \lambda_1^{(0)} <
  0.69$ for $\alpha >0$. The corresponding average value for $M_s =
  10^{16}~{\rm GeV}$ is $\langle \lambda_1^{(0)} \rangle = 0.31$, and
  for $M_s = 10^{19}~{\rm GeV}$ is $\langle \lambda_1^{(0)} \rangle =
  0.32$.

\item In Fig.~\ref{sm++_f6} we display the sensitivity of the RG
  evolution with $M_{Z''}$. For large values of $|\alpha|$ there is no
  variation in the contour regions. For $\alpha \agt -0.05$ and
  $\alpha \alt 0.06$ there are some small variances.  This small
  differences show the effect of the initial conditions
  $\lambda_2^{(0)}$ and $\lambda_3^{(0)}$ (both depending directly on
  $M_{Z''}$) on the evolution of the system.

\item We have verified that there is no significant variation of the
  SM$^{++}$ vacuum stability regions within the $m_H$ uncertainty.  An
  example for  $\alpha = 0.06$ and $M_{Z''} = 4.5~{\rm TeV}$ is given
  in Fig.~\ref{sm++_f7}.

\item In Fig.~\ref{sm++_f8} we display the variation of our
  results with $g'_1(M_s)$. It is clearly seen that for $0.232 < g'_1(M_s)
  < 1.000$ the dependence on $g_1'$ seems to be fairly
  weak.  The stability of SM$^{++}$ vacuum is then nearly
  independent of the $Z''$ branching fractions~\cite{Anchordoqui:2012wt}.

\item The low energy effective theory discussed in this paper requires
  a high level of fine tuning, which is satisfyingly resolved by
  applying the anthropic landscape of string
  theory~\cite{Bousso:2000xa,Susskind:2003kw,Douglas:2003um}. Alternatively, 
  the fine tuning can be circumvented with a more complete broken SUSY
  framework.  Since in pure SUSY the vacuum is automatically stable,
  the stability analysis perforce involves the soft SUSY-breaking
  sector. Hence rather than simply searching for the Higgs
  self-coupling going negative in the ultraviolet, the stability
  analysis would involve finding the local and global minima of the
  effective potential in the multi-dimensional space of the
  soft-breaking sector~\cite{Claudson:1983et}. However, the Higgs mass
  range favored by recent LHC data may be indicative of high-scale
  SUSY breaking~\cite{Hall:2009nd}; perhaps near the high energy
  cutoff of the field theory, beyond which a string description
  becomes a necessity~\cite{Hebecker:2012qp}.

\end{itemize}

In summary, we have shown that SM$^{++}$ is a viable low energy
  effective theory, with a well-defined range of free parameters.

\section*{Acknowledgments}
L.A.A.\ and BV are supported by the U.S. National Science Foundation (NSF)
under CAREER Grant PHY-1053663. I.A.\ is supported in part by the
European Commission under the ERC Advanced Grant 226371 and the
contract PITN-GA-2009- 237920. H.G.\ and T.R.T.\ are supported by NSF
Grant PHY-0757959.  X.H.\ is supported in part by the National
Research Foundation of Korea grants 2005-009-3843, 2009-008-0372, and
2010-220-C00003. D.L.\ is partially supported by the Cluster of
Excellence "Origin and Structure of the Universe", in Munich. D.L.\
and T.R.T.\ thank the Theory Department of CERN for its hospitality.
 L.A.A.\ and H.G.\ thank the Galileo Galilei
Institute for Theoretical Physics for the hospitality and the INFN for
partial support during the completion of this work.
Any opinions, findings, and conclusions or recommendations expressed
in this material are those of the authors and do not necessarily
reflect the views of the National Science Foundation.

\end{document}